\documentclass[epsf,usegraphicx]{mn2e}

\title[Monitoring the spin up in RX J0806+15]{Monitoring the spin up
in RX J0806+15}

\author[Hakala et al]
{Pasi Hakala$^{1}$, Gavin Ramsay$^{2}$, Kristiina Byckling$^{1}$\\
$^{1}$Observatory, University of Helsinki, PO Box 14, 
FIN-00014 University of Helsinki, Finland\\ 
$^{2}$Mullard Space Science Lab, University College London,
Holmbury St. Mary, Dorking, Surrey, RH5 6NT, UK}

\begin{document}
\outer\def\gtae {$\buildrel {\lower3pt\hbox{$>$}} \over 
{\lower2pt\hbox{$\sim$}} $}
\outer\def\ltae {$\buildrel {\lower3pt\hbox{$<$}} \over 
{\lower2pt\hbox{$\sim$}} $}
\newcommand{\ergscm} {ergs s$^{-1}$ cm$^{-2}$}
\newcommand{\ergss} {ergs s$^{-1}$}
\newcommand{\ergsd} {ergs s$^{-1}$ $d^{2}_{100}$}
\newcommand{\pcmsq} {cm$^{-2}$}
\newcommand{\ros} {\sl ROSAT}
\newcommand{\exo} {\sl EXOSAT}
\def\rchi{{${\chi}_{\nu}^{2}$}}
\newcommand{\Msun} {$M_{\odot}$}
\newcommand{\Mwd} {$M_{wd}$}
\def\Mdot{\hbox{$\dot M$}}
\def\mdot{\hbox{$\dot m$}}

\maketitle

\begin{abstract}

RX J0806+15 shows a prominent intensity variation on a period of 321.5
sec. This has widely been interpreted as the binary orbital period,
although this remains controversial. We have been monitoring the
precise period of RX J0806+15 for a number of years. By measuring the
rate of change we can help distinguish between competing physical
models. New observations obtained between Nov 2003 and Feb 2004 show
that the period decrease already reported by Hakala et al (2003) and
Strohmayer (2003) is continuing. We discuss how reliably we can
determine the period of RX J0806+15 using our technique and evaluate
the current models which have been proposed to account for the
observational properties of this source.

\end{abstract}


\section{Introduction}

Recently three sources have been discovered which show very stable
intensity variations on timescales of less than $\sim$10 mins: RX
J0806+15 (321 sec, Ramsay, Hakala \& Cropper 2002, Israel et al 2002);
RX J1914+24 (569 sec, Cropper et al. 1998, Ramsay et al. 2000, 2002)
and ES Cet (620 sec, Warner \& Woudt 2002). Furthermore, these are the
only periods which have conclusively been detected in these
systems. As such these periods have generally been taken to reflect
the binary orbital period. Such short periods imply a very small
binary dimension indicating that both stellar components are extremely
compact. However, their nature remains controversial (see Cropper et
al 2003).

One of the best techniques to resolve their nature is to determine
their period extremely accurately at different epochs. For binaries in
which mass transfer is occurring the binary orbital period should
increase over time. In contrast, in the case of the unipolar inductor
(UI) model (Wu et al 2002) the orbital period is expected to decrease
over time. On the other hand, Norton, Haswell \& Wynn (2004) argue
that the observed period is the spin period of the accreting white
dwarf and that any period decrease is consistent with that seen in the
weakly magnetic cataclysmic variables, the intermediate polars: IPs.

In the case of RX J0806+15, Hakala et al (2003) and Strohmayer (2003)
have presented evidence that the 321 sec period is decreasing at a
rate consistent with that expected if the system was being driven
entirely by gravitational radiation (ie consistent with the UI
model). More recently, Strohmayer (2004) has found evidence that the
period of RX J1914+24 is also decreasing over time.

However, these results are controversial, with Woudt \& Warner (2003)
claiming that there is difficulty in correctly identifying the
appropriate peak in the power spectra at different epochs. Clearly, it
is essential to monitor the period of these systems. We have a
programme of optical observations of RX J0806+15 to do this. In this
short paper we present our latest observational results and relate
these to the results of Hakala et al (2003). We discuss the
implications of these results regarding the nature of this system.

\begin{figure*} 
\begin{center} 
\setlength{\unitlength}{1cm}
\begin{picture}(13,9) 
\put(14.0,-1.5){\includegraphics{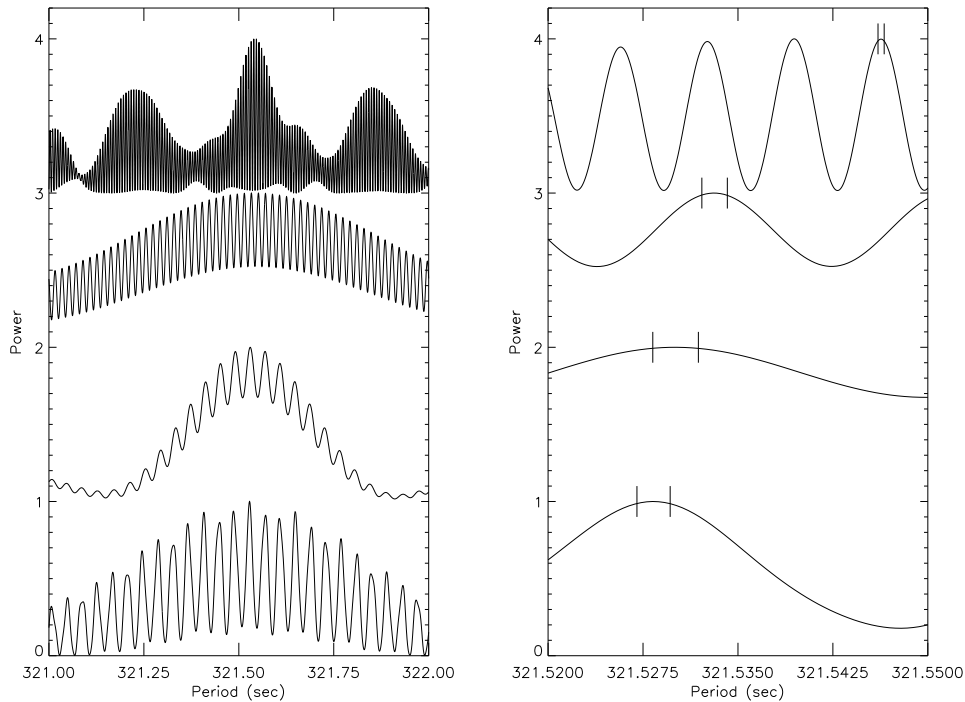}} 
\end{picture} 
\end{center} 
\caption{The
normalized (maximum equals 1.0) power spectra based on all the
observations available to us so far. The plots (from top to bottom,
shifted by 1.0 in Y-direction and in chronological order) are $ROSAT$
data, VLT 2001-NOT 2002 data, NOT 2003 data (all presented in Hakala
et al 2003) and INT 2003-NOT 2004 data. The short vertical lines
around the preferred peaks indicate the $\pm$ 3$\sigma$ error limits
for the best periods.}  
\label{spectrum} 
\end{figure*}

\section{Observations}

In order to measure possible period changes in RXJ0806+15, we have
obtained data from the Nordic Optical Telescope (NOT) and also the
Isaac Newton Telescope (INT), both located in La Palma. Table 1 shows
the observational log. The observations made using the INT were made
using the Wide Field Camera and were part of a project to detect
objects varying on short timescales (Ramsay \& Hakala in prep). White
light exposures were 30 sec in duration, but the readout time was of
the order of 40 sec implying a poor efficiency and an effective time
resolution of only 70 sec. In contrast, the observations made using
the NOT were dedicated to RX J0806+15. These observations were carried
out using ALFOSC in imaging mode, again in white light.  The
approximate time resolution was 20 sec (15 sec integration time) and
we selected only a small sub-window (approx 200x200 pixel) for fast
readout. In all our observations, frames were bias corrected and
flat-fielded in the usual manner. Star B (Ramsay, Hakala \& Cropper
2002) was used as the comparison. The date of the mid-point of the
exposures were heliocentric corrected.

\begin{table}
\begin{center}
\begin{tabular}{lrr}
\hline
Telescope & Dates & Duration \\
\hline
INT & 2003-11-29 & 2hr 6min\\
NOT & 2004-01-18 & 8hr 4min\\
NOT & 2004-01-19 & 7hr 42min (with 4hr 12min gap)\\
NOT & 2004-02-17 & 4hr 46min\\
\hline
\end{tabular}
\end{center}
\caption{The observation log of our INT 2003 and NOT 2004 observations.}
\label{obs}
\end{table}

\section{Period analysis}

\begin{figure} 
\begin{center} 
\setlength{\unitlength}{1cm}
\begin{picture}(8,5) 
\put(-1.0,-6.2){\includegraphics{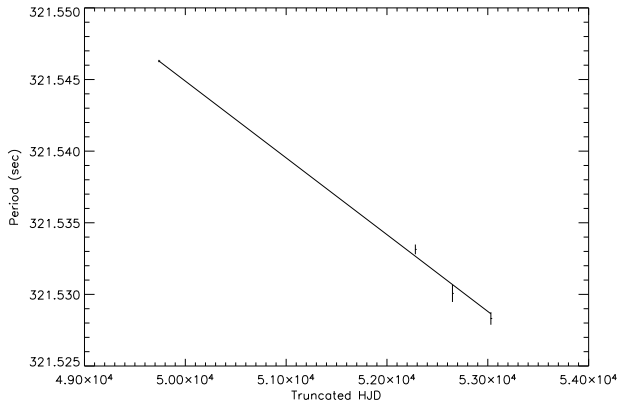}} 
\end{picture} 
\end{center} 
\caption{The period change of RX J0806+15 over time: the y-error bar 
reflects the error on the period.} 
\label{pdot} 
\end{figure}

We used the Lomb-Scargle power spectrum to search for the best fit
period of the combined INT and NOT light curve. The best fit period is
321.52832 sec and the resulting power spectrum is shown in Figure
\ref{spectrum} (lower curves). To determine the error on this period
we simulated 1000 datasets using original time bins together with a
sinusoidal equal amplitude variation with noise added at the observed
level. We find an error of 0.00044 sec. We show how this period
relates to the previous observations in Figure \ref{pdot}. This latest
period measurement confirms that the period is shortening (spinning
up) over time.

Hakala et al (2003) used 3 subsets of data to determine the change in
period of RX J0806+15. They found a period decrease at a rate of
3.14$\times10^{-16}$ Hz/s or 6.11$\times10^{-16}$ Hz/s depending on
which period they chose as the `real' period determined from {\sl
ROSAT} data. Based on this new period determination, we can rule out
the shorter of the two {\sl ROSAT} periods (321.5393 sec), (assuming
the orbital period is spinning up at a regular rate). Taking the
longer of the {\ros} periods and the other period measurements given
in Hakala et al (2003) with that here, we find that the system is
spinning up at a rate of 6.00$\times10^{-16}$ Hz/s ($\pm
1.0\times10^{-17}$ Hz/s).

\begin{table}
\begin{center}
\begin{tabular}{lrr}
\hline
Dataset & mid HJD & Period \\
\hline
ROSAT  & 2449738.010 & 321.54629 $\pm$ 0.00008 sec \\
VLT+NOT 2001-02 & 2452281.416 & 321.53314 $\pm$ 0.00034 sec  \\
NOT 2003 & 2452649.529 & 321.53007 $\pm$ 0.00060 sec \\
INT+NOT 2003-2004 & 2453030.395 & 321.52832 $\pm$ 0.00044 sec \\
\hline
\end{tabular}
\end{center}
\caption{The periods resulting from our analysis together with the errors from
Monte Carlo simulations.}
\label{pertable}
\end{table}

\section{Refined error analysis of the ROSAT data}

Our method of determining the periods has been criticised by Woudt \&
Warner (2003). While they agree with our error estimate for each power
peak they contend that we cannot be certain that we have identified the 
correct peak. We now address this concern using two
different techniques.

First we took the {\ros} data (barycentrically corrected) and binned
it into 10 sec bins.  We then folded it on the shorter of the {\ros}
periods (321.5393 sec, the highest peak in {\ros} power spectrum) to
obtain an average light curve shape (with 30 phase bins). To produce
synthetic data with the exactly same pulse shape and counting
statistics, we used the same {\ros} time points and for each time
point created a count rate using Poisson statistics with the mean
count rate for that particular phase interpolated from the phase
binned folded light curve. We then created 10000 simulated datasets
and ran the Lomb-Scargle analysis on these datasets.  In 84 \% of
cases, the highest power peak is the true period. In the remaining 16
\% of cases the highest peak is one of the neighbouring (alias) peaks,
which have equal probability.  Therefore, the peak corresponding to
the `true' period is {\sl never} lower than the second highest of all
peaks in the power spectrum.  In order to check our analysis against
using a `wrong period' for building up the {\ros} pulse shape (used as
a basis for our simulations) we repeated the simulations using
321.5463 sec (the second highest peak) period for folding instead. The
results from this test were identical to the earlier simulation.

The simulations described above only account for photon noise in the
data. If one assumes that there could be secondary effects in the
{\ros} data (like strong red noise or phase jitter due to the emission
region moving on the primary surface), then these could potentially
degrade our simulations. In order to account for these effects we have
also performed another type of (limited bootstrap) simulation. We
first take a period (321.5463 sec) and proceed by computing the
orbital phase for each of the original data-points (as in the first
approach). However, instead of trying to define any mean pulse shape,
we now use the actual {\ros} data-points to generate synthetic light
curves. We define a `phase correlation length' (pcl) to be the maximum
offset in orbital phase within which we can shuffle data-points. We
then take each of the time points and select the flux value for that
point from a pool of data-points that includes all the data with
orbital phase within the $\pm$1 pcl interval.

The main difficulty of this approach is the choice of the pcl value.
Firstly, one can take random pairs of {\ros} data-points and see how
the mean correlation between the pairs depends on their offset in
binary phase. In case of significant correlation (red noise or phase
jitter), one would expect a break in the mean correlation vs. phase
offset diagram at the value corresponding to the correct correlation
length. However, there is only a very small effect in such plot at pcl
= 0.04-0.05. Secondly, one can run the simulations at different pcl
values and measure when the peak distribution in the periodogram
starts to change significantly. This happens when the used pcl value
is larger than the real correlation length in the data (at pcl $\sim$
0.025 -0.03). Using simulations with pcl=0.025, we find that in 78.2\%
of the cases the highest peak is the true period. In 20.3\% of the
cases the true period is next to the highest peak and only in 1.5\% of
the cases the true period is lower than the second highest peak. These
results support our findings from our first (Monte Carlo) simulations.

Our simulations imply that the highest peak in the {\ros} data
(321.5393 sec) is either the true period or {\sl next to} the true
peak (with 98.5\% confidence). This alone would still allow 321.5393,
321.5463 or 321.5321 sec to be the true period. However, the measured
two highest peaks in actual {\ros} data are 321.5393 and 321.5463 sec,
which (based on our simulations) rules out the 321.5321 sec period (at
the 98.5\% level). However, we also have additional information from
our three optical period measurements.  If we use only these, we can
then estimate what the period would have been at the time of the
{\ros} observations. This exercise implies 321.54968 $\pm$ 0.0019
sec. Only one of the {\ros} periods, namely 321.5463 sec, is within
these limits.

Using our simulations we can also compute the error 
directly for the  {\ros} period. The error for any single peak in
the {\ros} power spectrum is significantly reduced, and is found to
be 0.00008 sec instead of 0.00040 sec, quoted earlier by Burwitz \&
Reinsch (2001) and Hakala et al. (2003). The earlier error estimates
were computed directly from the power spectrum assuming Cash
statistics, whilst here the error is defined as a standard deviation
of the best period from the Monte Carlo simulations.

Our results show that the aliasing problem for the {\ros} data is not
as serious as claimed by Woudt \& Warner (2003).  Their conclusion was
based on using their ES Cet {\sl optical} data to mimic the {\ros}
data sampling. Although they can mimic the {\sl sampling} of data,
their optical data cannot be used to mimic the X-ray modulation shape
nor the counting statistics, which probably resulted in an
overestimation of the sampling problem. The on-off type X-ray
modulation is very different in shape when compared to the
quasi-sinusoidal optical modulation. This implies that the phase
information of the X-ray data is more accurate, which in turn implies
that much more optical data would be required to match the `period
resolving power' of the X-ray power spectrum.

Woudt \& Warner (2003) also criticize the period determination for the
first of our optical datasets (Hakala et al. 2003). They claim that
the aliasing problem prevents us from measuring the true
period. However, the analysis of the aliasing structure reveals that
the possible periods for that dataset are 321.515, 321.534 and 321.552
sec. If these values are compared against all the other period
measurements (especially the two latest observations), it is
immediately clear that only one of the periods, namely 321.534s, is
possible.

%\begin{figure}
%\begin{center}
%\setlength{\unitlength}{1cm}
%\begin{picture}(8,5.5)
%\put(-1.5,-7){\special{psfile=psdsimu.ps hscale=50 vscale=50}}
%\end{picture}
%\end{center}
%\caption{The {\ros} power spectrum near the best period (solid line, offset
%in Y-direction by 50)) together with one of the power spectra resulting from our 
%Monte Carlo simulated datasets overplotted (dashed line).}
%\label{syntpsd}
%\end{figure}

\section{Discussion and Conclusions}

Our observations confirm the results of Hakala et al (2003) and
Strohmayer (2003) who find that RX J0806+15 is spinning up. We find a
spin-up rate of 6.0$\times10^{-16}$ Hz/s -- consistent with that
determined using the `long' {\ros} period as found by Hakala et al
(2003). If we have correctly identified the 321 sec period as the
orbital period, the fact that the orbital period of RX J0806+15 is
spinning up puts a question mark against accretion driven models. It
does not rule them out since this could be just a secular change as
seen in other binaries rather than a long term change. On the other
hand it is spinning up at a rate consistent with that of being driven
purely by gravitational radiation, which is what is predicted by the
UI model (Wu et al 2002). If the 321 sec is identified with that of
the spin period of the primary star (as suggested in the IP
interpretation, Norton et al 2004), then the spin up is consistent
with that seen in other IPs. We now discuss the relative merits of
these two models in relation to RX J0806+15.

The applicability of the UI model to RX J0806+15 rests on the 321 sec
period being correctly identified as the binary orbital period. If a
second, longer, period can conclusively be identified then that would
have a strong claim to be the binary orbital period and the UI model
would therefore not be relevant to this source. On the other hand this
model can account for all the observational properties, including the
rate of spin up. It has been claimed that since Hydrogen is blended
with Helium in the weak emission lines of RX J0806+15 (Israel et al
2002) this argues against all double degenerate models, including the
UI model, (Norton et al 2004, Reinsch, Burwitz \& Schwarz 2004). This
is because the minimum orbital period for a degenerate Hydrogen-rich
companion is $\sim$30 min (Rappaport, Joss \& Webbink 1982). However,
even for Helium-rich stars, a measurable amount of Hydrogen is still
present in white dwarfs (eg Friedrich et al 2000).  We conclude
therefore that the presence of weak Hydrogen emission lines in the
spectrum of RX J0806+15 does not rule out the double degenerate models
and that the UI model is still a viable model for RX J0806+15. Indeed,
it can account for all the observational properties of this system.

We now consider the IP model developed by Norton et al (2004). In this
model the 321 sec period represents the spin period of the stream
accreting white dwarf. The fact that a second longer period has not
been detected is explained by the system being a face-on binary system
and hence there is no observational signature of the binary orbital
motion (Norton et al 2004). Photometric observations extending into
the IR by Reinsch et al (2004) argue against a typical main sequence
secondary star as is typical in IPs. Norton et al (2004) suggest that
in RX J0806+15 the secondary is a brown dwarf. They also note that a
double degenerate IP model is possible. However, they argue that this
is unlikely because of the presence of Hydrogen in its optical
spectra. Again, we do not consider this is a valid reason for
excluding this model. The IP interpretation has been criticised
because in contrast to all other IPs, there are no strong emission
lines in either RX J0806+15 or RX J1914+24. Norton et al (2004) argue
that most of the line emission originates near the base of the
accretion column and is obscured by the stream having a high optical
depth. However, the strongly magnetic systems, the polars, which have
a virtually identical stream geometry to that proposed by Norton et al
(2004), show stream emission extending relatively far from the
accreting white dwarf and not just from the base of the accretion
column. We conclude that the absence of strong emission lines in these
systems is a drawback for this model.

The nature of both RX J0806+15 and RX J1914+24 remains uncertain.  Our
current set of period measurements span less than ten years. Even if
both these systems appear to be spinning up at a rate expected from
general relativity, it is not yet {\sl certain} that this is the cause
for the spin up. However, we believe that the UI model remains, at
this stage, the model which best accounts for the observational
properties of RX J0806+15. Perhaps the definitive observations will be
phase-resolved spectroscopy - if the weak emission lines are modulated
on a 321 sec timescale, then that would be strong evidence that this
period is indeed the binary orbital period.

\section{Acknowledgments}

PJH is an Academy of Finland research fellow.  Based on observations
made with the Nordic Optical Telescope, La Palma.  The data were obtained
with ALFOSC, which is owned by the Instituto de Astrofisica de Andalucia
(IAA). Observations were also made using the INT Wide Field
Camera. We gratefully acknowledge the support of the staff at both
Observatories.

{}

\end{document}